\documentclass[nofootinbib,amsfont,amssymb,amsmath, showpacs,balancelastpage, nofootinbib,aps,pra,twocolumn]{revtex4-2}
\usepackage{CJKutf8}

\usepackage{soul}
\usepackage{comment}

\newcommand{\newtext}[1]{\textcolor{blue}{#1}}

\usepackage{amsthm}

\newtheorem{lemma}{Lemma}
\theoremstyle{definition}
\newtheorem{defn}{Definition}

\usepackage[utf8]{inputenc}
\usepackage{amsmath,amssymb,amsfonts,stmaryrd}
\usepackage{physics}
\usepackage{dsfont}
\usepackage{graphicx}
\usepackage{xcolor}
\usepackage{graphicx}
\usepackage{cancel}
\usepackage{natbib}
\usepackage{color}
\usepackage{tikz}
\usetikzlibrary{matrix}
\usepackage{mathrsfs}
\usepackage{relsize}
\usepackage{multirow}
\usepackage[linktocpage]{hyperref}
\usepackage[all]{xy}
\hypersetup{colorlinks=true,citecolor=blue,linkcolor=blue, urlcolor=blue, breaklinks=true}
\usepackage[scr=boondox]{mathalpha}
\usepackage{diagbox} 

\usepackage{float}

\usepackage{bm} 
\usepackage{environ}
\usepackage{url}
\usepackage[margin=0.75in]{geometry}

\usepackage{mathtools}

\newcommand{\Z}{{\mathbb Z}}

\def\U{\mathrm{U}}

\newcommand{\SO}{\text{SO}}
\newcommand{\SU}{\text{SU}}

\NewEnviron{eqs}{%
\begin{equation}\begin{split}
    \BODY
\end{split}\end{equation}}

\begin{document}
\title{In search of diabolical critical points}
\author{Naren Manjunath}
\affiliation{Perimeter Institute for Theoretical Physics, Waterloo, Ontario N2L 2Y5, Canada}

\author{Dominic V. Else}
\affiliation{Perimeter Institute for Theoretical Physics, Waterloo, Ontario N2L 2Y5, Canada}
\begin{abstract}
A phase transition is an example of a ``topological defect'' in the space of parameters of a quantum or classical many-body systems. In this paper, we consider phase diagram topological defects of higher codimension. These have the property that equilibrium states undergo some kind of non-trivial winding as one moves around the defect. We show that such topological defects exist even in classical statistical mechanical systems, and describe their general structure in this context. We then introduce the term ``diabolical critical point'' (DCP), which is a higher-codimension analog of a continuous phase transition, with the proximate phases of matter replaced by the non-trivial winding of the proximate equilibrium states. We propose conditions under which a system can have a stable DCP. We also discuss some examples of stable DCPs in (1+1)-dimensional quantum systems.

\end{abstract}
\maketitle

\section{Introduction}

Topological defects such as vortices, skyrmions and their generalizations have long played a key role in shaping our understanding of phases of matter. Some famous applications of topological defects in classical physics include the theory of ordered media \cite{mermin1979}, the Kosterlitz-Thouless transition \cite{Kosterlitz_1973}, and melting transitions \cite{Halperin1979}. Equally well-known are their applications to topological phases of quantum systems, in which topological defects can have quantized charges under additional symmetries.

The above applications all involve an order parameter that varies in space or time, and the topological defect corresponds to some spacetime configuration of the order parameter with non-trivial homotopy invariants. In this paper, we instead consider \textit{phase diagram topological defects} in classical and quantum many-body systems. We consider systems in which the Hamiltonian contains some number of parameters that can be varied, leading to a phase diagram. In other words, we consider \textit{families} of equilibrium states. Suppose the overall parameter space is $\mathbb{R}^n$ for some $n$. Then there may be a subset of parameters, $\mathcal{S} \subseteq \mathbb{R}^n$, at which something ``singular'' happens: for example, if the surrounding family of states is gapped, then $\mathcal{S}$ can correspond to surfaces where the system becomes gapless. We call this configuration in the phase diagram a ``phase diagram topological defect'' if the surrounding family of equilibrium states parameterized by $\mathbb{R}^n \setminus \mathcal{S}$ admits some kind of non-trivial ``winding''(or a suitable generalization). In such cases, just as with a vortex core for instance, there \emph{must} be singularity somewhere in $\mathcal{S}$ for topological reasons. The defining property of a non-trivial parameterized family is that the ground states undergo a non-trivial discrete process, such as a charge pumping or a permutation in the space of ground states, as we perform a winding around the phase diagram defect. Examples of such families have been discussed in several recent works in the context of generalizations of the Thouless pump \cite{Kapustin2020Thouless,Cordova2020anomcoupling,Shiozaki2022adiabatic}, the higher Berry curvature \cite{Kapustin2020HBC,Hsin2020berry,Wen2021HBC,Qi2023MPS,shiozaki2025HBC-MPS}, the breaking of continuous symmetries \cite{else2021goldstone,Debray2023SB,manjunath2024anomalous,bose2025csl}, and the space of conformal boundary conditions \cite{wen2025conformalBC,choi2025conformalBC}. While these works focus on quantum many-body systems, non-trivial winding is also possible in \emph{classical} statistical mechanics systems~\cite{Sun2025,sun2025strain}, as we will see in this paper.

\begin{figure}
    \centering
    \includegraphics[width=0.8\linewidth]{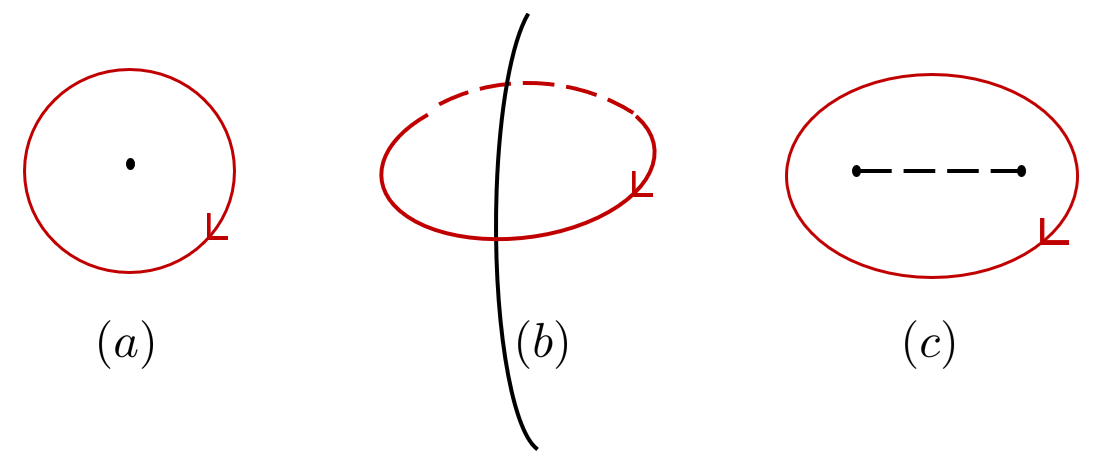}
    \caption{(a) The simplest case of a diabolical critical point (DCP) occurs for a parameter space $\mathbb{R}^N$ containing a family over $S^{N-1}$. In this case the DCP has codimension $N$ and is a single point. Here we show the case $N=2$ (the DCP is marked in black, and a loop indicating the $S^1$ family is shown in red) (b) We can include additional parameters that change the parameter space to $\mathbb{R}^n$ for $n>N$ but there is still a DCP in parameter space if the singular surface of the defect has codimension $N$. (c) If the singular surface has codimension less than $N$, it is not a DCP.}
    \label{fig:DCP-examples}
\end{figure}

Just as with more familiar topological defects in space, phase diagram topological defects come with a concept of ``dimensionality''. For example, topological defects in three spatial dimensions can be line defects (e.g.\ vortices) or point defects (e.g.\ hedgehogs). One can make a similar classification for phase diagram topological defects; however, an important point is that since one can always add additional parameters, it is only the \emph{codimension} of the defect that has a robust meaning. For example, a point defect in a two-dimensional parameter space will generally become a line defect in three-dimensional parameter space when an additional parameter is added -- but the codimension of 2 is unchanged. See Fig.~\ref{fig:DCP-examples} for an illustration.

It is helpful to think of phase diagram topological defects as a generalization of the concept of a ``phase transition''. Indeed, phase diagram topological defects of codimension 1 are precisely phase boundaries. In that case the equivalent of ``non-trivial'' winding is that the phase transition separates two phases that are distinct in some way (for example, they have different patterns of spontaneous symmetry breaking, or different topological order). In this paper, we will consider phase diagram topological defects of higher codimension.

A very simple example of a phase diagram topological defect is as follows. Suppose we have a quantum spin in a magnetic field $\vec{B}$ with Hamiltonian $H = -\vec{B} \cdot \vec{S}$. It can be described by a phase diagram in $\mathbb{R}^3$, corresponding to the values of $\vec{B}$. The ground states $\ket{\psi}_{\vec{B}}$ are such that the spin component along $\vec{B}$ equals $+s$. For fixed $|\vec{B}| \neq 0$, this gives a family of gapped ground states parameterized by the 2-sphere $S^2$. The family can be associated with a quantized topological invariant, namely the Chern number of the Berry connection around this monopole, which for a spin $s$ particle equals the integer $2s$. The ground state at $B = 0$ consists of $(2s+1)$ degenerate states and preserves the full $\SO(3)$ spin rotation symmetry. If $s$ is a half-integer, this symmetry is furthermore anomalous (it acts projectively). The monopole at the origin of parameter space is an example of a phase diagram topological defect of codimension 3.

The concepts of topological family and phase diagram topological defect can in fact describe a variety of other interesting phenomena that can occur even in classical statistical mechanical systems. Consider, for example, systems with a spontaneously broken Ising symmetry, in which there are two symmetry-breaking equilibrium states. If we consider a loop in parameter space, we can trace how the symmetry-breaking ground states evolve as we move around the loop. Once we go all around the loop, one of two things can have happened. Either the symmetry-breaking equilibrium states come back to themselves, or they get interchanged. In the latter case, this corresponds to a non-trivial family of ground states with a discrete topological invariant (the interchange property). This gives rise to a phase diagram topological defect of codimension 1.
One main result of this paper is to give the general theory of topological families associated with SSB in classical systems, see Section \ref{sec:General-Classical} below. This includes systems with \emph{continuous} spontaneously broken symmetries, where there are examples of phase diagram topological defects of higher codimension.

The second main issue which we will discuss in this paper is: what exactly is the nature of the singular core of a phase diagram topological defect? The case we view as the ``most interesting'' is where one has what we refer to as a \emph{diabolical critical point (DCP)}, which is a generalization of the concept of a continuous phase transition. A diabolical critical point is a particular kind of phase diagram topological defect of codimension $N$, where the singular core has no ``thickness'', i.e. it is precisely a surface of codimension $N$ (for example, in the case $N=2$, with a two-dimensional parameter space, the core is a single point rather than having nonzero size), and moreover the expectation values of local observables evolve continuously as one approaches the core. See Fig.~\ref{fig:DCP-examples} for an example. In the case $N=1$, this reduces to the usual notion of a continuous phase transition. As with usual continuous phase transitions, one expects that the IR physics when sitting on the core in parameter space must be described by a scale-invariant field theory (i.e.\ an RG fixed point), with the non-trivial family it is proximate to imprinted in some way on the critical fluctuations.

In the example of a quantum spin in a magnetic field with $H = - \vec{B} \cdot \vec{S}$, there is indeed a single gapless codimension 3 point at the origin, where the different spin states become degenerate, and the system has full $\SO(3)$ spin rotation symmetry. A DCP is an analog of this example in higher space dimensions. Some examples of DCPs have previously been pointed out in Ref.~\cite{Hsin2020berry}.

A perennial question of study in the theory of phase transitions is to find and characterize quantum field theories which can describe continuous phase transitions between two given phases of matter. In this paper, we take some steps towards addressing the analogous question for DCPs. Specifically, we put forward a hypothesis about the general properties that a DCP enclosed in any given family must satisfy. We cannot prove these properties, but they hold in all the examples we consider, and they lead to a way to construct specific examples of DCPs. The hypothesis is as follows. Consider a family of ground states with symmetry $G$ parameterized by the $N-1$-sphere $S^{N-1}$. Then we believe that the key property of a DCP contained within the family is that it must have an emergent symmetry $G$ and a subgroup $H$ such that $G/H = S^{N-1}$. The idea is that moving away from the DCP should correspond to adding terms to the Hamiltonian that break the emergent symmetry $G$ down to $H$. 

A DCP is not the only possibility for the singular core of a phase diagram topological defect enclosed by a topological family over an $N-1$-sphere. Another possibility is that the set $\mathcal{S}$ of singular points in parameter space becomes a surface of codimension less than $N$, with the possibility of a first-order transition as one crosses the surface, or of a critical surface. The latter scenario has previously been referred to as ``unnecessary criticality'' \cite{Bi2019Adventure,Prakash2024Multi,Prakash2024pumps}, although in the context of phase diagram topological defects it is in fact \emph{necessary} that there be some kind of singularity in the phase diagram. The former possibility will become apparent from the examples we discuss.

Let us mention some related prior work, which primarily deals with quantum many-body systems. The notion of a DCP with an emergent symmetry generalizes the known result that the usual codimension-1 phase boundaries between distinct symmetry-protected topological phases are associated with an emergent $\Z_2$ or $\Z_2^T$ symmetry which is anomalous \cite{Bultinck2017,Tsui2015}. To our knowledge, the notion of diabolical point was first spelled out in a quantum field theory context in Ref.~\cite{Hsin2020berry}. Ref.~\cite{Tantiv2023Pivot} studied conditions on the lattice for a $\U(1)$ symmetry to emerge at a gapless point in a phase diagram, although the gapless points they study are not strictly DCPs since there are multiple distinct phases surrounding them. High-symmetry points along the loops defining a topological family were recently studied in Ref.~\cite{jones2025pivot}.

The rest of the paper is organized as follows. In Sec.~\ref{sec:classical} we discuss classical topological families coming from SSB; some technical computations for this section are left for Appendix~\ref{app:S2Family}. In Sec.~\ref{sec:DCP-criterion} we discuss the properties of DCPs with some examples. To fully state our criterion for DCPs in the quantum case we need to use the `compatibility' framework recently developed in Ref.~\cite{manjunath2024anomalous}. We conclude in Sec.~\ref{sec:Disc}.

\section{Classical examples from spontaneous symmetry breaking}\label{sec:classical}

The main point we wish to make in this section is that topological families and phase diagram topological defects are not limited to quantum systems at zero temperature. In this section we will discuss topological families of classical statistical mechanical systems in which each member of the family exhibits spontaneous symmetry breaking (SSB). Some examples of this phenomenon were recently studied in Refs.~\cite{Sun2025,sun2025strain}.

\subsection{Ising-SSB family}\label{sec:IsingSSB}

Consider a system with a $\mathbb{Z}_2$ symmetry that is spontaneously broken. Then there are two symmetry-breaking equilibrium states $\rho_1$ and $\rho_2$. Now suppose in the phase diagram, we have a family parameterized by the circle $S^1$, such that at each point on the circle, the $\mathbb{Z}_2$ symmetry is spontaneously broken. If we move infinitesimally along the circle, the symmetry-breaking equilibrium states $\rho_1$ and $\rho_2$ will be slightly perturbed to $\rho_1'$ and $\rho_2'$, but it should still be possible to unambiguously identify $\rho_1$ as the deformation of $\rho_1'$ and $\rho_2$ as the deformation of $\rho_2'$. Suppose, however, that we follow this through all the way around the circle. Then it can happen that $\rho_1$ and $\rho_2$ get \emph{exchanged}. This will correspond to a non-trivial ``winding'' in the space of $\mathbb{Z}_2$ SSB equilibrium states. (For a similar effect in the context of quantum field theories, see Ref.~\cite{Sharon2020}).

For example, consider a classical thermal system with two real scalar fields written as $\vec{\phi} = (\phi_1, \phi_2)$, with a $\Z_2$ symmetric Hamiltonian. 
If we express $\vec{\phi}$ as $\vec{\phi} = R (\cos \theta, \sin \theta)$, the $\Z_2$ symmetry corresponds to invariance under $\theta \rightarrow \theta + \pi$. For simplicity, let us work at zero temperature, so that the equilibrium state is simply the state that minimizes the energy. Furthermore, let us assume that the that the energy takes the form 
\begin{equation}\label{eq:H-IsingSSB}
    E = \int d^d x [ (|\vec{\triangledown} \theta|^2 + a \cos 2 \theta + b \sin 2 \theta + V(R) ]
\end{equation}
where $a,b \in \mathbb{R}$. We will assume that the potential $V(R)$ is chosen such that the Hamiltonian is minimized at $R = 1$, and the only variable in the problem is $\theta$. Let $\lambda = \arctan \frac{-b}{a}$. Note that for fixed $a,b$ with $a^2 + b^2 >0$ the minima of $\mathcal{H}$ occur at two values $\theta^*, \theta^* + \pi$, where $ \theta^* = \frac{\pi - \lambda}{2}$. In this case, the ground states form a space $\Omega$ consisting of two points which are permuted by the $\Z_2$ symmetry. Therefore, we obtain a family of doubly degenerate ground states parameterized by $\lambda \in S^1$. 

The system selects a minimum by spontaneously breaking the $\Z_2$ symmetry. Now consider starting in some specified ground state and slowly changing $\lambda$; for convenience we can keep $a^2 + b^2 > 0$ fixed. The value of $\theta^*$ changes with $\lambda$. In particular, winding $\lambda$ around the origin by $2\pi$ takes the system from one minimum $\theta^*$ into the other minimum $\theta^* + \pi$, and we find that the initial ground states have become switched.

Let us distinguish this from a trivial $\Z_2$-SSB family, which consists of two degenerate ground states for every choice of an $S^1$ order parameter, but cycling the order parameter does not permute the equilibrium states. This can be realized by any constant family, in which the equilibrium states are completely independent of the $S^1$ parameter.

\begin{figure}
    \centering
    \includegraphics[width=0.35\textwidth]{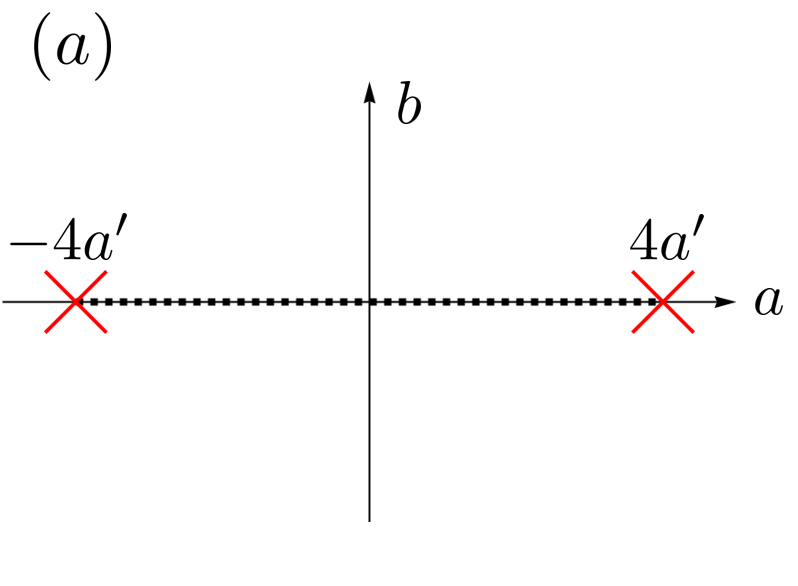} \includegraphics[width=0.3\textwidth]{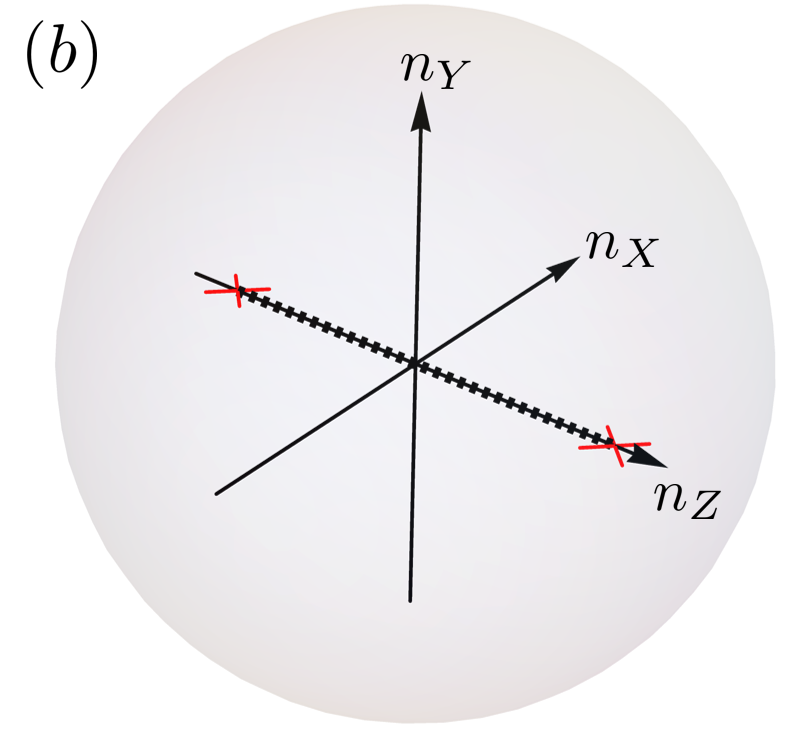}
    \includegraphics[width=0.28\textwidth]{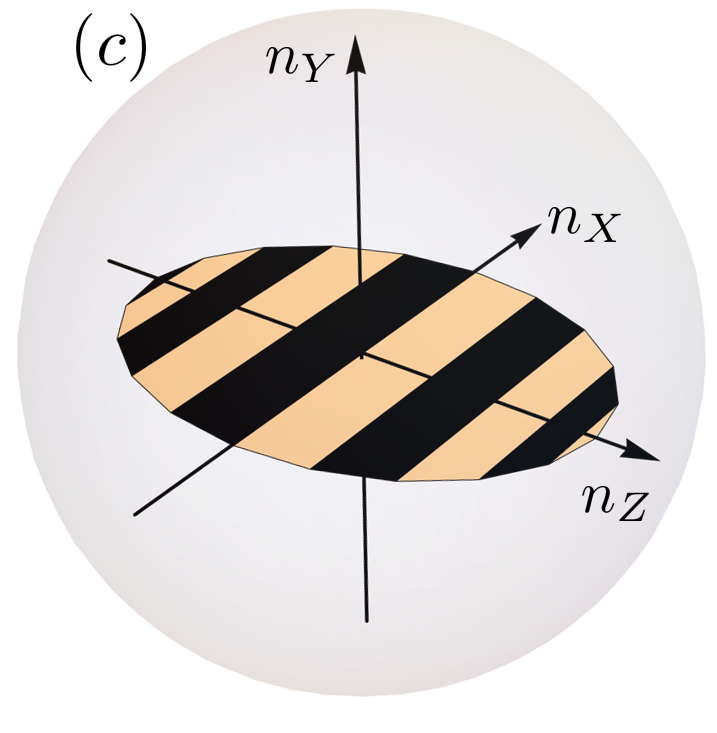}
    \caption{Stability of DCP at the origin for (a) the Ising SSB family with the perturbation in Eq.~\eqref{eq:IsingSSBPerturbation}, and (b) the SSB family over $S^2$ with the perturbation in Eq.~\eqref{eq:S2FamPerturbation}.  The DCP splits into a first order line (dashed) terminating in a pair of critical points (red `X' marks). In (c) there are two perturbations corresponding to Eq.~\eqref{eq:XZPerturbation} which turn the DCP into a first-order ellipse.}
    \label{fig:IsingFamily}
\end{figure}

In the non-trivial example, the origin $a=b=0$ is a singular point in the parameter space spanned by $(a,b)$. In the absence of any higher-order cosine terms in $\theta$, every other point in the $(a,b)$ phase diagram is doubly degenerate and spontaneously breaks the $\Z_2$ symmetry. We now test the stability of this singular point to generic perturbations of the Hamiltonian. For example, we can add the symmetry-allowed term $a'\cos (4 \theta)$ to the theory, so the $\theta$-dependent terms become
\begin{align}\label{eq:IsingSSBPerturbation}
    a \cos 2\theta + b \sin 2\theta + a' \cos 4\theta. 
\end{align}
If we plot the minima $\theta^*$ in the $(a,b)$ plane for fixed $a'>0$ (see Fig.~\ref{fig:IsingFamily}(a)), we find that there are two minima everywhere except on the segment $b = 0, |a| \le 4 a'$, where there are four minima. There is a first order transition upon crossing this segment: $\theta^* (a,b = 0^+) = -\theta^* (a,b = 0^-)$. As a result, the phase diagram topological defect now becomes a `diabolical locus' (this terminology was used in Ref.~\cite{Hsin2020berry}), corresponding to the first-order line $\{(x,0), |x| \le 4 a'\}$. There is still a well-defined SSB family defined on paths that encircle the first-order line. The line itself is not an example of the diabolical critical points (DCP) defined in the introduction, since the topological defect under consideration has codimension 2, and therefore a DCP would be a single point in the 2-dimensional $(a,b)$ parameter space.

\subsection{General mathematical formulation of classical SSB families}\label{sec:General-Classical}
We can generalize the above example to give a general theory of topological invariants for families of SSB states, as follows. Consider a family of Hamiltonians $\mathcal{H}(\lambda)$, $\lambda \in \Lambda$. Suppose that for each choice of $\lambda$, $\mathcal{H}(\lambda)$ preserves a symmetry $G$, but the equilibrium state spontaneously breaks $G$ down to a subgroup $H$. The symmetry-breaking equilibrium states are parameterized by a space $\Omega := G/H$ carrying a $G$ action. $\Omega$ is a group manifold if $H$ is a normal subgroup, but we will not make this assumption about $\Omega$ below. In fact, in Appendix~\ref{app:S3} we discuss a simple example in which $\Omega$ is not a group.

We will assume that a nontrivial SSB family over $\Lambda$ is defined by a nontrivial fiber bundle $\Omega \rightarrow \mathcal{E} \rightarrow \Lambda$. Although we cannot rigorously prove this assumption in general, it seems intuitively clear. Here $\mathcal{E}$ physically represents the space of all possible minima of $\mathcal{H}(\lambda)$ (for all possible values of $\lambda \in \Lambda$). The idea is that for each value of the parameter $\lambda \in \Lambda$, there is a space of SSB ground states in one-to-one correspondence with $\Lambda_{\mathrm{GS}}$. However, for different parameter values $\lambda_1, \lambda_2$, the SSB ground states are not identical states, and hence there is no \emph{canonical} way to identify the space of SSB ground states for the two parameter values. (This is the essence of what it means to have a fiber bundle.)

In fact, we can be a bit more specific. A general fiber bundle admits a \emph{structure group} $\hat{G}$, which is a subgroup of the group of all self-homeomorphisms of the fiber, such the transition functions can be chosen to be valued in $\hat{G}$. We can argue what the structure group for the fiber bundles under consideration should be, as follows. Physically, one expects that there should be some kind of quasi-adiabatic evolution operator that induces a map from the space of ground states at one value $\lambda$ of the parameters to the space of ground states at a different value $\lambda'$ of the parameters, given a continuous path from $\lambda$ to $\lambda'$. (One can rigorously construct such a map in gapped quantum many-body systems, which would be relevant for the case of discrete SSB; however we expect that some kind of map will exist more generally). This defines a ``connection'' on the fiber bundle. Given that, by assumption, the Hamiltonian is invariant under the $G$ symmetry for all values of the parameters, it follows that this connection will be equivariant with respect to the action of the $G$ symmetry on states. In the absence of too many pathologies, this will imply (see Appendix \ref{appendix:structure_group}) that the fiber bundle admits structure group
\begin{equation}
\label{eq:structure_group_1}
\hat{G} = \mathrm{Aut}(\Omega)_G.
\end{equation}
Here $\mathrm{Aut}(\Omega)_G$ is the group of homeomorphisms from $\Omega$ to itself that commute with the $G$ action on $\Omega$. One can show (see Appendix \ref{appendix:structure_group_equivalence}) that this is equivalent to
\begin{equation}
\label{eq:structure_group_2}
    \hat{G} = N_G(H)/H,
\end{equation}
where
\begin{equation}
    N_G(H) = \{ g \in G : gH = Hg \}
\end{equation}
is the normalizer of $H$ in $G$.

It is a standard fact from the theory of fiber bundles that, once the fiber and the $\hat{G}$ action in it have been specified, fiber bundles with structure group $\hat{G}$ are in one-to-one correspondence with \emph{principal} $\hat{G}$ bundles. Thus we obtain our main result:
\begin{quote}
    When a symmetry $G$ is spontaneously broken to $H$, SSB families over a parameter space $\Lambda$ are classified by principal $\hat{G}$-bundles over $\Lambda$, where $\hat{G}$ is defined by \eqref{eq:structure_group_2}.
\end{quote}

In the special case when $H$ is trivial, $\hat{G}= N_G(H) = G$. This occurs for the Ising SSB family, where $G = \Z_2, H = \Z_1$; the SSB families are classified by principal $\Z_2$ bundles over $S^1$, which are equivalently maps from $S^1 \rightarrow B\Z_2$ (where $B\Z_2$ is the classifying space of $\Z_2$). The classification is therefore given by $\pi_1(B\Z_2) = \pi_0(\Z_2) = \Z_2$, matching the conclusion from the previous section.

\subsection{SSB family with a continuous ground state manifold}
In order to illustrate the above general considerations, we discuss a more complicated example in which we obtain a non-trivial SSB family over the 2-sphere $S^2$. Consider a system in which the order parameter is a two-component complex field $z = \binom{z_1}{z_2}, z_1, z_2 \in \mathbb{C}$. Again, we work at zero temperature for simplicity. The energy is 

\begin{equation}\label{eq:Higher-SSB}
    E =  - z^{\dagger} ~ (\hat{n} \cdot \vec{\sigma})  ~z + g |z|^4 + \dots
\end{equation}
Observe that the system has a $G=\mathrm{U}(1)$ symmetry corresponding to shifting $z \to e^{i\theta } z$. We pick $g$ so that the free energy is minimized when $|z| = \sqrt{|z_1|^2 + |z_2|^2} = 1$. Let $z_j = |z_j| e^{i \theta_j}, j = 1,2$. An explicit calculation of the minimum $z_*$ for each $\hat{n} \in S^2$ is given in Appendix~\ref{app:Minimize-classicalSSB}; the result is
\begin{align}\label{eq:z-minimum}
    |z^*_1| = \cos \beta ; \quad |z^*_2| = \sin \beta; \quad
    \theta^*_1 - \theta^*_2  = \tan^{-1}(n_Y/n_X)
\end{align}
for $\beta \in [0,\pi/2]$ satisfying $\cos 2\beta = n_Z$. For each $\vec{n}$ the different minima form an $S^1$ manifold $\Omega = S^1$ and are related by the $\U(1)$ symmetry mentioned above, which shifts $\theta_1^*, \theta_2^*$ equally. Therefore for each $\vec{n}$ the system must spontaneously break the $\U(1)$ symmetry down to a trivial group $H = \Z_1$.

In the presence of a spatially varying $\vec{n}$ vector, both $z_1^*$ and $z_2^*$ will vary. Now consider inserting a skyrmion of $\vec{n}$. To be concrete, consider a cross-section of the system in the two-dimensional $xy$ plane, and suppose the skyrmion is centered at the origin, with an $(r,\phi)$ polar coordinate representation
\begin{align}
    n_X &= \sin 2\gamma \cos \phi, n_Y = \sin 2\gamma \sin \phi, n_Z = \cos 2\gamma; \nonumber \\ \sin \gamma &= e^{-r/2}.
\end{align}
Here $\hat{n} = \hat{z}$ as $r \rightarrow \infty$, so that $|z_1^*| = 1, |z_2^*|=0$. This means that the space of equilibrium states is the \emph{same} (not just isomorphic) for each point on a large circle with $r \to \infty$, which allows us to unambigously define a ``winding number'' in the space of ground states as one moves around the circle. Indeed, as we go around the large circle, $\theta_1^*$ winds by $2\pi$ and returns to its original value. Therefore, a skyrmion of $\hat{n}$ induces a non-trivial winding in the manifold of degenerate states.

Mathematically, the family is given by a fiber bundle with fiber $\Omega = S^1 \simeq \SO(2)$ and base space $\Lambda = S^2$. The total space $\mathcal{E}$ corresponds to the space of possible values of $z^*$, which is the 3-sphere $S^3$. These spaces fit into the fibration $\Omega \rightarrow \mathcal{E} \rightarrow \Lambda$. Since $H = \Z_1$, the structure group of the bundle is $\hat{G} = G = \U(1)$. The classification of SSB families in this case is given by $\U(1)$ principal bundles over $\Lambda = S^2$, or equivalently by maps from $S^2 \rightarrow B\U(1)$. These maps are classified by $\pi_2(B\U(1)) = \pi_1(\U(1)) = \Z$, which agrees with the above interpretation that a skyrmion with unit topological charge induces windings in the $\Omega = S^1$ space of ground states with winding number $m \in \Z$. The specific example above has $m=1$.

If we ignore all other symmetry-allowed terms in the energy, the parameter space has a singular point ($n_X = n_Y = n_Z = 0$). As in the Ising SSB problem, we can study the stability of this singular point to additional perturbations. Since the free energy is a dipole term in $z$, a natural choice for these perturbations is the quadrupole term $V(z) = \sum_{ij} n_{ij} (z^{\dagger} \sigma^i z) (z^{\dagger} \sigma^j z)$, where $\sigma^i, i = X, Y, Z$ are the Pauli matrices, and $n_{ij}$ are real numbers. Let us consider a specific choice in which the only nonzero quadrupole term has the coefficient $n_{ZZ} >0$. We then have
\begin{equation}
    E = - z^{\dagger} (\vec{n} \cdot \vec{\sigma}) z + n_{ZZ} (|z_1|^2-|z_2|^2)^2 + \dots 
\end{equation}
where $n_X^2 + n_Z^2 = 1$, and the additional terms represented by $\dots$ enforce $|z_1|^2+|z_2|^2 = 1$. Defining $\chi = |z_1|^2-|z_2|^2$, this becomes
\begin{align}\label{eq:S2FamPerturbation}
  E &= n_{ZZ} \left(\chi - \frac{n_Z}{2 n_{ZZ}}\right)^2 - (n_X + i n_Y) z_1^* z_2 \nonumber \\ & \quad- (n_X - i n_Y) z_2^* z_1 + \dots  
\end{align}
Whenever $n_X = n_Y = 0$ and $|n_Z/2 n_{ZZ}| <1$, the minimum is achieved by picking $\chi = n_Z/2 n_{ZZ}$. This fixes $|z^*_1|$ and $|z^*_2|$, so the only remaining freedom is to rotate the arguments $\theta^*_1$ and $\theta^*_2$. Now let us turn on $n_X, n_Y$ but make them very small, so that $|z_1^*|, |z_2^*|$ remain the same. Then the minimum is achieved when
\begin{equation}
    \theta^*_1 - \theta^*_2 = \tan^{-1}(n_Y/n_X).
\end{equation}
In particular, going from $(n_X, n_Y, n_Z)$ to $(-n_X, -n_Y, n_Z)$ changes $\theta^*_1 - \theta^*_2$ by $\pi$. Therefore we have a first-order line defined by $n_X = n_Y = 0, - 2 n_{ZZ} \le n_Z \le 2 n_{ZZ}$ (see Fig.~\ref{fig:IsingFamily}(b)). For $|n_Z/2n_{ZZ}|>1$, we must set either $z_1^* = 0$ or $z_2^* = 0$ depending on the sign of $n_Z$, and there is still a DCP.

We could in fact add two separate quadrupole perturbations, for example
\begin{align}\label{eq:XZPerturbation}
    E &= - z^{\dagger} (\vec{n} \cdot \vec{\sigma}) z + n_{ZZ} (|z_1|^2-|z_2|^2)^2 \nonumber \\ &\quad+ n_{XX} (z_1^* z_2 + z_1 z_2^*)^2+ \dots 
\end{align}
This case is studied in Appendix~\ref{app:S2Family}. We find the phase diagram in Fig.~\ref{fig:IsingFamily}(c). There is now a first-order \textit{plane}, charcterized by the ellipse $n_Y = 0, (n_X/2n_{XX})^2 + (n_Z/2n_{ZZ})^2 \le 1$. The minimum jumps discontinuously as we pass through this region from $n_Y \rightarrow 0^+$ to $n_Y \rightarrow 0^-$. 

\subsection{More examples}

Below we briefly consider some additional examples in which the residual symmetry $H$ after SSB is non-trivial, implying that the structure group $\hat{G}$ is not isomorphic to the unbroken symmetry $G$.

\subsubsection{Heisenberg ferromagnet}

One such example is that of the `Heisenberg ferromagnet', in which a spin system whose free energy has a $G = \SO(3)$ spin rotation symmetry subsequently undergoes SSB down to $H = \SO(2)$ by picking a uniform direction for the average magnetization. Here $N_G(H) = \U(1) \rtimes \Z_2$; the $\U(1)$ subgroup of $N_G(H)$ consists of arbitrary rotations about the magnetization axis, while the additional $\Z_2$ subgroup is generated by $\pi$ rotations about any axis perpendicular to the magnetization. The structure group in this case is $\hat{G} = \Z_2$. In the identification of $\hat{G} = \mathrm{Aut}_G(\Lambda)$ (in this case $\Lambda = S^2$ corresponds to the choice of magnetization direction), $\hat{G}$ is generated by the inversion map which sends each point on the sphere to its antipode. A non-trivial topological family exists because there is a symmetry preserving automorphism, i.e. the inversion map, which is non-trivial (not in $H$). 

Note that principal $\hat{G}$ bundles over $S^1$ have a non-trivial $\Z_2$ classification. This means that we can consider evolving the ferromagnetic state around a parameter space $\Lambda = S^1$, and in the case of a non-trivial family, the magnetization flips with each winding around $\Lambda$. Note, however, that there is no element of $G = \SO(3)$ which sends the magnetization $\vec{n} \rightarrow -\vec{n}$ for \textit{every} choice of $\vec{n} \in S^2$. Therefore, this parameter space topological defect does not correspond to a $G$ symmetry defect, in contrast to the examples above.

\subsubsection{`Dirac point' in parameter space}

An example similar to the one we now discuss was presented in Ref.~\cite{Sun2025} as an analog of a `Dirac point' in Laudau theory. Consider systems with $G = \text{O}(2) = \U(1) \rtimes \Z_2^T$ symmetry, realized for instance by a two component easy-plane order parameter which can be continuously rotated or reflected in the $x-y$ plane. The reflection can be naturally realized by a time-reversal operation, although the description of the SSB family is not affected by instead considering unitary reflections. The system undergoes SSB by picking a direction in the $x-y$ plane, which breaks the $G$ symmetry down to $H = \Z_2^T$ (for instance, if the chosen direction is $+\hat{x}$, then the equilibrium state still preserves reflection symmetry about the $x$ axis).

The parameters of the SSB family are therefore $\Lambda = S^1, G = \U(1) \rtimes \Z_2^T, N_G(H) = \Z_2\times\Z_2^T, \hat{G} = \Z_2$. In particular, the SSB families are classified by principal $
\Z_2$ bundles over the circle, which gives a classification $\pi_0(\Z_2) = \Z_2$. The interpretation is somewhat similar to the Ising SSB family discussed previously: there are two equilibrium states that are symmetric under reflections about a chosen axis, and the non-trivial SSB family switches them upon winding the $S^1$ parameter. 

\section{Diabolical critical points: Properties and Examples}\label{sec:DCP-criterion}

As mentioned in the introduction, a diabolical critical point (DCP) is a generalization of a continuous phase transition in which there is a singular surface of codimension $N$ in parameter space that is enclosed by a non-trivial family over $S^{N-1}$. The non-trivial family must be imprinted in some way on the critical fluctuations of the DCP. In this section, we will make some hypotheses about the ways in which we expect this to manifest.

Specifically, we will state some properties we expect a (stable) DCP of codimension $N$ to have.
We will consider both the classical SSB case and the quantum case reviewed below. We do not have a formal justification for these properties, but we will attempt to give some heuristic arguments for why these conditions are natural to expect. The properties are:

\begin{itemize}
    \item A DCP of codimension $N$ should correspond to an RG fixed point with exactly $N$ relevant operators $\phi_1, \cdots, \phi_N$ (it will follow from the properties described below that they all have the same scaling dimension).
    \item It should have an emergent continuous symmetry $\hat{G}$  with some action on $\vec{\phi} = (\phi_1, \cdots, \phi_N)$ by $O(N)$ matrices. The corresponding $\hat{G}$ action on the unit sphere should be transitive. (That is, any two points on the unit sphere can be related by the action of an element of $\hat{G}$).
    \item The microscopic symmetry $G$ should map into $\hat{G}$ via a homomorphism $\varphi : G \to \hat{G}$ whose image contains only elements with trivial $O(N)$ action.
\end{itemize}

The fact that there should be exactly $N$ relevant operators is clear, given that we want the DCP to have codimension $N$ in parameter space.
The statement about the emergent symmetry is much less obvious. Nevertheless, we can give some justifications for why this statement should hold, as follows.
Firstly, we note that these properties do hold in the small number of examples of DCPs that we can construct explicitly (see Section~\ref{sec:StableDCP} below).

Secondly, the statement about the emergent symmetry can be viewed as a generalization of the statement that phase transitions between symmetry-protected topological phases necessarily have an emergent $\mathbb{Z}_2$ or $\mathbb{Z}_2^T$ symmetry \cite{Tsui2015,Bultinck2017} (which corresponds to the case $N=1$).

Finally, what we will argue is that \emph{if} one has a CFT or other RG fixed point satisfying these properties, then one can obtain a non-trivial topological family over $S^{N-1}$ upon adding the relevant operators, provided that certain compatibility conditions are satisfied as we describe below.

For example, suppose we have a microscopic symmetry $G = \mathbb{Z}_2$, and suppose that we consider a DCP with two relevant operators $\phi_1$ and $\phi_2$ and emergent symmetry $\hat{G} = \mathrm{SO}(2)$, which contains $G$ as a subgroup and acts on the relevant operators twice as fast as the standard action of $\mathrm{SO}(2)$ on $\mathbb{R}^2$. Then, adding the relevant operators to the Hamiltonian explicitly breaks the symmetry back down to the $\mathbb{Z}_2$ subgroup of $\mathrm{SO}(2)$ corresponding to $G$. If we assume that adding the relevant operators causes the system to flow to a state in which the $\mathbb{Z}_2$ symmetry is spontaneously broken, then one can show that the corresponding family of SSB states parameterized by $S^1$ is described by the fiber bundle
\begin{equation}
    \mathbb{Z}_2 \to \mathrm{SO}(2) \to S^1.
\end{equation}
(recall that such fiber bundles classify the families over $S^1$, see Section \ref{sec:General-Classical}).

One can generalize this result as follows. Let $G$ be the microscopic symmetry, and suppose the general conditions described above are satisfied. Moreover, suppose that upon adding the relevant operators, the residual explicit symmetry $G' \leq \hat{G}$ is spontaneously broken down to a subgroup $H' \leq G'$. Then the microscopic symmetry is spontaneously broken down to $H = H' \cap G$. We assume that $G/H = G'/H'$ (otherwise the spontaneous symmetry breaking would not be stable to perturbations respecting the microscopic symmetry). Then one can show that the topological invariant of the family of SSB states parameterized by $S^{N-1}$ is fully determined by $\hat{G}$, $H'$ and the action of $\hat{G}$ on the space of relevant operators. 

Below, we will show how in the case of zero-temperature quantum families, the family over $S^{N-1}$ that one obtains upon adding the relevant operators will also be related to the properties of the emergent symmetry $\hat{G}$.

\subsection{Compatibility properties: quantum case}

\subsubsection{Review: Parameterized families of quantum many-body states}

In this section, we will minimally review parameterized families of quantum many-body states, which have recently been studied in several works as cited in the introduction. For further details about the compatibility conditions in particular, see Ref.~\cite{manjunath2024anomalous}.

A parameterized family of gapped quantum many-body ground states is specified by two pieces of data: a topological space $\Lambda$ and a symmetry group $G$ (the microscopic symmetry). The family corresponds to a collection of gapped ground states $\ket{\psi_{\lambda}}$ which all preserve the $G$ symmetry, and in fact can be adiabatically connected to each other without breaking $G$. The non-trivial property of such a family is that topological defects of the order parameter living in $\Lambda$ carry discrete quantum numbers under $G$.

To simplify the discussion we will focus on \textit{invertible} families in the remainder of this section. A gapped ground state $\ket{\Psi}$ is \textit{invertible} if there exists another state $\ket{\Psi^{-1}}$ such that $\ket{\Psi} \otimes \ket{\Psi^{-1}}$ can be adiabatically connected to a trivial product state. The state is called non-invertible if it is not invertible. We sometimes use the term `$G$-invertible states' to refer to invertible states with symmetry $G$. A family $\{\ket{\psi}_{\lambda}\}$ is called invertible (non-invertible) if each $\ket{\psi_{\lambda}}$ is invertible (non-invertible). 

A \textit{constant} family over $G$ is one in which each $\ket{\psi_{\lambda}}$ is the same state. A family is called a \textit{pump} if each $\ket{\psi}_{\lambda}$ can be connected to a product state without breaking $G$ symmetry. Invertible families in space dimension $d$ with symmetry $G$ are classified by an Abelian group, similarly to invertible states. The classification of invertible families canonically splits into a product of constant families and pumps:
    \begin{align}
\label{eq:family_product}
& \{\text{Invertible $G$-families over $\Lambda$}\}_d \nonumber \\= &\{\text{$G-$inv} \}_d \times \{ \text{$G$-pumps over $\Lambda$} \}_d
\end{align}
where $\{\text{$G$-inv} \}_d$ denotes the classification of invertible states with symmetry $G$ (i.e. the constant families) and $\{ \text{$G$-pumps over $\Lambda$} \}_d$ denotes the classification of pumps. Therefore any family $f$ can be expressed as a pair $(c(f),\pi(f))$ where $c(f)$ is the constant part and $\pi(f)$ is the pump part.

\subsubsection{The compatibility result for DCPs}

Consider some $G$ invariant family $f$ over $S^{N-1}$ (where $G$ is the UV symmetry of the family) with a DCP of codimension $N$ at the center of the family. The DCP has IR symmetry $\hat{G}$, and there is a homomorphism $\varphi: G \rightarrow \hat{G}$ with $G_{IR} \leq \hat{G}$ the image of $\varphi$. Then, the assumption of a transitive action of $\hat{G}$ on $S^{N-1}$ means that we can apply the framework of Refs.~\cite{else2021goldstone,manjunath2024anomalous}, in which for a family over $S^{N-1}$ with a transitive action of $\hat{G}$, there should exist a subgroup $\hat{H}$ with $\hat{G}/\hat{H} = S^{N-1}$. $\hat{H}$ is physically the subgroup of $\hat{G}$ that leaves a given point on the sphere fixed. $G_{IR}$ is a subgroup of $H$ which leaves the whole of $S^{N-1}$ fixed.\footnote{A notational comment: what we call $(\hat{G},\hat{H},G_{IR},\Lambda = \hat{G}/\hat{H})$ is denoted in Ref.~\cite{manjunath2024anomalous} as $(G,H,H_0,\Lambda = G/H)$.}  
    
Ref.~\cite{manjunath2024anomalous} further gives a compatibility condition which applies in the quantum case. If we restrict to invertible families, there is a $G_{IR}$-symmetric pump $p$ which should be compatible with the $\hat{G}$-anomaly at the gapless point, in the sense of Ref.~\cite{manjunath2024anomalous}. Moreover, we must have $\varphi^*(p) = \pi(f)$ where $\pi(f)$ is the pump part of the family $f$ and $\varphi^*$ denotes the pullback with respect to $\varphi$. In words, the homomorphism $\varphi$ can be used to view the $G_{IR}$ pump as a $G$ pump.

\subsection{Examples of phase diagrams with stable DCPs}\label{sec:StableDCP}
In the quantum case, it is actually possible to show that some well-known parameterized families in (1+1)D have stable DCPs.

\subsubsection{Thouless pump}
Ref.~\cite{Hsin2020berry} previously analyzed the low-energy theory of the canonical Thouless pump in 1d. The gapless point is described by a compact boson CFT, and it was shown in Ref.~\cite{Hsin2020berry} that there is a regime around the non-interacting limit of this CFT with only two relevant operators that generate the Thouless pump. This implies that the gapless point is in fact a DCP of codimension 2. Below we briefly review this result.

We define $2\pi$-periodic dual fields $\theta, 
\phi$, and the gapless point in the phase diagram has two $\U(1)$ symmetries corresponding to invariance under shifts of $\theta$ and $\phi$. There is a mixed anomaly between the two $\U(1)$ symmetries. We assume that the shift symmetry of $\theta$ is the $\U(1)$ symmetry preserved in the Thouless pump. We can perturb away from the gapless point by inserting vertex operators into the theory. We use the conventions of Ref.~\cite{Hsin2020berry} in which at radius $R$, the vertex operator $e^{i n \theta + i m \phi}$ has conformal dimensions
\begin{equation}
    (h,\bar{h}) = \left( \frac{1}{2}\left(\frac{n}{R} + \frac{mR}{2}\right)^2,\frac{1}{2}\left(\frac{n}{R} - \frac{mR}{2}\right)^2\right).
\end{equation}
An operator with conformal dimensions $(h,h)$ is relevant if $h<1$. By symmetry, we are allowed to add perturbations of the form $\cos m\phi, \sin m\phi$, which have conformal dimension $h = \bar{h} = m^2 R^2/8$. Note that if $R^2>8$, all operators of this form are irrelevant, while if $2 < R^2 < 8$, the only symmetry-allowed and relevant operators are $\cos \phi, \sin \phi$, and they have the same scaling dimension. 

For this range of $R$ we can verify that the system satisfies all the postulated properties of a DCP at the beginning of Sec.~\ref{sec:DCP-criterion}. (i) There are exactly $N=2$ relevant operators with the same scaling dimension. (ii) The emergent continuous symmetry of the gapless point is $\hat{G} =\U(1)_{\theta} \times \U(1)_{\phi}$. It contains the microscopic symmetry (which is isomorphic to $\U(1)_{\theta}$) as a subgroup. The $\U(1)_{\phi}$ subgroup of $\hat{G}$ acts on the unit circle, defined by the relevant operators, transitively via $\SO(2)$ matrices. (iii) The image of the microscopic symmetry of the Thouless pump, which is $\U(1)_{\theta}$, has a trivial action on the relevant operators. (iv) Finally, the fact that we get a Thouless pump upon adding the relevant operators implies, according to the general compatibility relation of Ref.~\cite{manjunath2024anomalous}, that there must be a mixed anomaly beteween the two $\mathrm{U}(1)$'s, which is indeed the case. 

When $R^2 < 2$, the operators $\cos 2\phi, \sin 2\phi$ also become relevant. In this case we expect that the DCP is turned into a first-order line, as argued in Ref.~\cite{Hsin2020berry}.

\subsubsection{Family with higher Chern number}

Another example is the 1d family over $S^3$ characterized by the higher Chern number, which is an invariant classified by $H^3(S^3,\Z)$. We will focus on the family with higher Chern number 1, which has attracted considerable interest recently. One possibility is that the family is realized by starting with an $\SU(2)$ symmetric state, with anomaly given by a Wess-Zumino-Witten (WZW) term at level $k \in \Z$ in the action. The case $k=1$ can be described by the free-fermion limit of the Thouless pump ($R=2$ in the CFT discussed above), where in addition to $\cos \phi, \sin \phi$ we are allowed to add the operators $\cos \theta, \sin \theta$, and these operators all have the same scaling dimension.

We now study whether the family over $S^3$ can have a stable DCP. We use the fact that the IR fixed point of the gapless point associated with $S^3$ family in the absence of any $\SU(2)$-breaking perturbations is given by the $\SU(2)$ WZW CFT at level $k=1$, which is reviewed for example in Ref.~\cite{Fradkin-book}. We wish to find all possible relevant symmetry-breaking perturbations to the CFT. 
The primary fields in this theory are indexed by operators in the spin-$j$ representation of $\SU(2)$, where $j = 0,1/2,1,\dots, k/2$. The scaling dimension $\Delta_j$ of the spin-$j$ primary is given by
\begin{equation}
    \Delta_j = \frac{2j(j+1)}{k+2}.
\end{equation}
This operator is relevant if $\Delta_j < 2$. Now, if we restrict to the theory with $k=1$, the only primaries correspond to $j = 0,1/2$. The $j=0$ primary is just the identity operator. The spin-1/2 primaries transform as a complex 2-dimensional representation of $\SU(2)$, therefore there are four of them and they all have the same scaling dimension $\Delta_{1/2} = 1/2$. When $k=1$ these operators are also relevant, and generate the desired family over $S^3$ by completely breaking the $\SU(2)$ symmetry. That we obtain the correct family can be checked by working out the compatibility relation between the $\SU(2)$ anomaly and the higher Chern number which is a $\Z$ invariant for the family over $S^3$. In this case, the anomaly index is also an integer invariant given by the level of the WZW term, i.e. $k=1$. The compatibility condition was worked out mathematically in Ref.~\cite{manjunath2024anomalous}; it simply states that the higher Chern number of the family we obtain should equal the anomaly coefficient $k=1$.

Since there are no other primary operators in the theory, we can directly conclude that there is a stable DCP for the family with higher Chern number $k=1$. Let us again check each condition postulated previously. (i) This DCP has codimension 4, since there are exactly 4 relevant operators which gap out the theory, and they have the same scaling dimension. (ii) In this case the emergent symmetry of the DCP is $\hat{G} = \SU(2)$, while the microscopic symmetry of the topological family is trivial: $G = \Z_1$, therefore $G_{IR}$ is also trivial. The parameter space of the family is $\SU(2) \cong S^3$, and $\hat{G}$ acts transitively on this space via a homomorphism from $\SU(2)$ to $\SO(4)$. (iii) Since $G$ is trivial, the third condition is trivially satisfied. Furthermore, the compatibility condition between the $\hat{G}$ anomaly and the surrounding family is also satisfied, as we noted above.    

In contrast, let us return to the classical Ising-SSB family discussed in Sec.~\ref{sec:classical}. In space dimension 2, the classical model can also be mapped to a \newtext{(1+1)d} compact boson CFT, and this can be used to study the stability of the DCP to symmetry-allowed perturbations. 
We use the same conventions for the fields in the CFT as above. But in this case, we wish to break the $\U(1)$ shift symmetry of $\theta$ down to $\Z_2$ by adding the operators $\cos 2\theta, \sin 2\theta$. The $\Z_2$ symmetry will then be spontaneously broken, realizing the Ising SSB family. These operators are relevant when $2/R^2 < 1$, i.e. $R> \sqrt{2}$. Observe that the operators $\cos 4\theta, \sin 4\theta$ only become relevant when $8/R^2<1$, i.e. when $R>2\sqrt{2}$. Therefore, when $\sqrt{2} < R < 2\sqrt{2}$, the DCP is stable to perturbations involving only the variable $\theta$. Note, however, that for this range of $R$, $R^2/8<1<R^2/2$, and so the operators $\cos \phi, \sin \phi$ are also relevant, while $\cos m \phi, \sin m \phi$ are irrelevant for all $m \ge 2$. (The same mathematical analysis appears in the rather different context of intrinsically gapless topological phases in Ref.~\cite{Thorngren2021igSPT}). Therefore it is necessary to study the competition between the $\cos 2\theta, \sin 2\theta$ terms and the perturbations $\cos \phi, \sin \phi$, which when dominant will proliferate vortices of $\theta$, restoring the original $\U(1)$ symmetry. We expect that the presence of the relevant $\cos \phi$ and $\sin \phi$ terms will generically destabilize the DCP.

\subsubsection{Dirac fermion in (3+1) dimensions}

A higher-dimensional example in a similar spirit to the (1+1)-dimensional cases above, also considered in Ref.~\cite{Hsin2020berry}, is the (3+1)0-dimensional Dirac fermion, with Lagrangian
\begin{equation}\label{eq:Dirac_3+1}
    \mathcal{L} = -i \overline{\Psi} \cancel{\partial} \Psi + m \overline{\Psi} (\cos \theta + i \sin \theta \sigma^y) \Psi 
\end{equation}
where $\Psi = (\psi_1, \psi_2)$ is a two-component Dirac spinor, $\overline{\Psi} = \Psi^{\dagger} \gamma^0 = \Psi^{\dagger} \sigma^z$, $m>0$ is a fixed real number, and $\theta \in S^1$ is a parameter. In the absence of the mass term, the system has two $\U(1)$ symmetries corresponding to separate charge conservation within each component. The mass term breaks this down to the diagonal $\U(1)$ subgroup corresponding to overall charge conservation. The symmetry broken by the mass term is the `axial' $\U(1)$ symmetry which acts on the Dirac spinor as $\Psi \rightarrow e^{i \alpha \sigma^z} \Psi$. 

This system realizes a topological family over $S^1$, in which winding the parameter $\theta$ pumps an integer quantum Hall (IQH) state. It is simplest to see this in momentum space. We can fix some momentum coordinate $k_z$, choosing $|k_z|<|m|$. Then, the system at $\theta = 0,\pi$ realizes a gapped (2+1)-dimensional Dirac fermion in the $x-y$ plane, with mass $k_z \pm m$. Therefore, tuning the parameter effectively changes the Chern number of this (2+1)-dimensional cross-section by one unit. This is analogous to  the case of the (1+1)-dimensional Thouless pump, in which a unit charge is pumped across a \textit{real-space} cross-section viewed at a fixed position $x$. Note that Ref.~\cite{Hsin2020berry} instead considered a family over $S^3$, which is obtained by breaking the $\SU(2)$ symmetry of the Dirac term down to nothing; the resulting family is a pump of an $\SU(2)$ quantum Hall state rather than an IQH pump as we have discussed here.

In any case, it is straightforward to check that all interaction terms are irrelevant by power-counting. As a result, the singular point $m=0$ of the above $S^1$ family forms a stable DCP of codimension 2. The DCP has a mixed anomaly of the $\U(1) \times \U(1)$ symmetry, which satisfies the compatibility condition of Ref.~\cite{manjunath2024anomalous}.

\section{Discussion}\label{sec:Disc}

Topological families of gapped quantum many-body states are now receiving substantial attention in the condensed matter and quantum field theory literature. One of the main results in this paper is to extend the notion of topological families to classical systems with spontaneous symmetry breaking. In the examples we study, each member of the family corresponds to a system with degenerate equilibrium states, and the system spontaneously breaks the symmetry. The signature of a non-trivial family is that performing a nontrivial loop in the parameter space leads to a permutation, or discrete winding, of the equilibrium states. We gave a mathematical characterization of such families in terms of fiber bundles.

We also have taken the first steps towards a systematic study of \emph{diabolical critical points} (DCPs). Classical or quantum criticality is one of the most profound emergent behaviors that many-body systems can exhibit, and DCPs are a fascinating new kind of criticality which is related to topological families rather than phase transitions. An important question is how the topological family gets imprinted on the critical fluctuations of the DCP. The conjectures that we have put forward in this paper about the properties that a DCP must satisfy provide a starting point for studying this question.

An immediate question emerging from this work is to find classical SSB families with stable DCPs: the conditions for this to occur are still not clear. Another is to find more examples of quantum systems with stable DCPs, particularly in two and higher dimensions. It would be interesting to approach this question by using the conformal bootstrap approach to study potential CFTs satisfying the properties that we have conjectured DCPs must satisfy, e.g.\ $N$ relevant operators related by some $O(N)$ action of some emergent symmetry.

\section{Acknowledgments}

We thank Abhishodh Prakash, Joseph Maciejko, Po-Shen Hsin, Marvin Qi, Xueda Wen, Ryan Lanzetta, and Zhen Bi for discussions. Research at Perimeter Institute is supported in part by the Government of Canada through the Department of Innovation, Science and Economic Development; and by the Province of Ontario through the Ministry of Colleges, Universities, Research Excellence and Security.

\appendix

\section{Computations for classical SSB family over $S^2$}\label{app:S2Family}
\subsection{Minimization of Eq.~\eqref{eq:Higher-SSB}}\label{app:Minimize-classicalSSB}
Let $z_j = |z_j|e^{i \theta_j}, j=1,2$. The terms defining the SSB family in Eq.~\eqref{eq:Higher-SSB} can be expanded as
\begin{align}
    F =& -2n_X |z_1| |z_2| \cos (\theta_1 - \theta_2) -  2n_Y |z_1| |z_2| \sin (\theta_1 - \theta_2) \nonumber \\ &- n_Z (|z_1|^2 - |z_2^2|)
\end{align}
where $n_X^2 + n_Y^2 + n_Z^2 = 1$ and, by assumption, $|z_1^2| + |z_2^2| = 1$. Taking $|z_1| = \cos \beta, |z_2| = \sin \beta$ for some $\beta \in [0,\pi/2]$, we get
\begin{widetext}
    \begin{align}
    F &= -n_X \sin 2\beta \cos (\theta_1 - \theta_2) - n_Y \sin 2\beta \sin (\theta_1 - \theta_2) - n_Z \cos 2\beta \\
    &= - n_Z \cos 2\beta - \sqrt{1-n_Z^2} \sin 2\beta \cos (\theta_1 - \theta_2 - \tan^{-1}(n_Y/n_X)). 
\end{align}
\end{widetext}

To minimize this expression we should pick $\theta^*_1 - \theta^*_2 = \tan^{-1}(n_Y/n_X)$. Then the quantity left to minimize is
\begin{align}
    F &= - n_Z \cos 2\beta - \sqrt{1-n_Z^2} \sin 2\beta \\
    &= - \cos (2 \beta - \tan^{-1}(1/n_Z)).
\end{align}
This expression is minimized when $\cos 2\beta = n_Z$. The minimum is therefore achieved when
\begin{align}
    |z^*_1| &= \cos \beta = \sqrt{\frac{1+n_z}{2}} \nonumber \\
    |z^*_2| &= \sin \beta = \sqrt{\frac{1-n_z}{2}} \nonumber \\
    \theta^*_1 - \theta^*_2 & = \tan^{-1}(n_Y/n_X)
\end{align}
and there is an overall $\U(1)$ freedom in selecting $\theta^*_1, \theta^*_2$, which is ultimately achieved by spontaneous symmetry breaking.

\subsection{Analysis of Eq.~\eqref{eq:XZPerturbation}}
With the same parameterization as above, and additionally setting $\delta = \theta_1 - \theta_2$, Eq.~\eqref{eq:XZPerturbation} becomes
\begin{widetext}
    \begin{align}
    F & = - n_Z \cos 2\beta + n_{ZZ} \cos^2 2\beta - n_X \sin 2\beta \cos \delta + n_{XX} \sin^2 2\beta \cos^2 \delta - n_Y \sin 2\beta \sin \delta  \\
    &= n_{ZZ} \left(\cos 2\beta - \frac{n_Z}{2n_{ZZ}}\right)^2 + n_{XX} \left(\sin 2\beta \cos \delta - \frac{n_X}{2n_{XX}}\right)^2 - n_Y \sin 2\beta \sin \delta .
\end{align}
\end{widetext}

First assume $n_Y = 0$. Let $a_Z = |n_Z/2n_{ZZ}|, a_X = |n_X/2n_{XX}|$. When $a_Z^2 + a_X^2 \le 1$, the minimum is achieved by setting both square terms to zero: 
\begin{equation}
    \cos 2\beta^* = \frac{n_Z}{2n_{ZZ}}; \quad \sin 2\beta^* \cos \delta^* = \frac{n_X}{2n_{XX}} .
\end{equation}
The surface $\{\vec{n}~|n_Y = 0, a_Z^2 + a_X^2 \le 1\}$ forms an ellipse in the $(n_X, n_Z)$ plane with axes of length $2n_{ZZ}$ and $2n_{XX}$. Note that since we have only fixed $\cos \delta^*$ there are two possible solutions for $\delta^*$, which are distinguished by the sign of $n_Y$. Therefore in this case, the minimum jumps discontinuously as we cross the ellipse, implying that we have a first-order \textit{plane}. 

\section{An example where $\Omega$ is not a group}\label{app:S3}

In the two previous examples, we were able to give both $\Omega$ and the total space of the associated fiber bundle the structure of a group. But this is not the case when $\Omega = G/H$ where $H$ is not a normal subgroup of $G$. A simple choice is $G = S_3 = \Z_3 \rtimes \Z_2$, and $H = \Z_2$. In this case, $\Omega$ is a set of three points with a $\Z_3$ permutation action. It is not a group: defining $r$ as the $\Z_3$ rotation and $\sigma$ as the reflection in $S_3$, the three points are given by the pairs $\{r^k, r^k \sigma\}$ for $k = 0,1,2$. Since $r$ is of order 3 and $r \sigma$ is of order 2, there is no consistent way to define $\{r, r \sigma\}$ as a group element. However, there is a well-defined $\Z_3$ action on $\Omega$, defined for example through left multiplication by $r$.

We can consider a family of these ground states over $S^1$, such that winding the order parameter leads to a $\Z_3$ permutation. The family is given by a fiber bundle $\Omega \rightarrow \mathcal{E} = S^1 \rightarrow \Lambda$. But to see the full structure, we can instead consider the fiber bundle $S_3 \rightarrow \text{O}(2) \rightarrow \Lambda$. This allows us to visualize the gapless point of the SSB family as a system with $\text{O}(2)$ symmetry. We can first explicitly break $\text{O}(2)$ down to $S_3$; then spontaneously breaking $S_3$ down to $\Z_2$ gives an SSB family with the desired permutation action on the ground states. 

\section{Structure group of fiber bundle with a $G$ action}
\label{appendix:structure_group}
First we recall the definition of structure group.
\begin{defn}
Let $\pi : E \to B$ be a fiber bundle with fiber $F$. Let $\hat{G}$ be a subgroup of the group of homeomorphisms from $F$ to $F$. Then we say that the fiber bundle admits structure group $\hat{G}$ if there exists a collection of open sets $\{ U_i \}$ that cover $B$, and local trivializations $\varphi_i : \pi^{-1}$ such that image of each of the transition functions $t_{ij} : (U_i \cap U_j) \to \mathrm{Aut}(F)$ is contained in $\hat{G}$, where $\mathrm{Aut}(F)$ is the group of homeomorphisms from $F$ to itself.
\end{defn}

Then we have
\begin{lemma}
Let $\pi : E \to B$ be a fiber bundle with fiber $F$. Let $G$ be a topological group, and let $* : G \times E \to E$ be a continuous group action on $E$ that preserves fibers. Choose some $b_0 \in B$, and choose some homeomorphism $\phi_0 : \pi^{-1}(b_0) \to F$. Then $*$ induces a $G$ action $*_0$ on $F$ such that 
\begin{equation}
    (b_0, g *_0 f) = (b_0, \phi_0(g * \phi_0^{-1}(f))).
\end{equation}
for all $f \in F$.

Now, suppose that it is possible to find a collection of open sets $\{ U_i : i \in I \}$ that cover $B$, and local trivializations $\varphi_i : \pi^{-1}(U
_i) \to U_i \times F$ which are equivariant functions with respect to the $G$ action $*$ on $\pi^{-1}(U_i)$ and the $G$ action $*_0$ on $U_i$ that is derived from the trivial action on $U_i \times F$ and the action $*_0$ on $F$. Then the fiber bundle admits structure group $\hat{G} = \mathrm{Aut}(F)_G$, where $\mathrm{Aut}(F)_G$ is the group of homeomorphisms from $F$ to itself that commute with the $G$ action.

\begin{proof}
The corresponding transition functions are determined by the compositions
\begin{multline}
    \hat{t}_{ij} : {(U_i \cap U_j)} \times F \to {(U_i \cap U_j)} \times F, \\ \hat{t}_{ij} = \varphi_j \circ \varphi_i^{-1}.
\end{multline}
The $G$-equivariance of $\varphi_{i,j}$ thus ensures the $G$-equivariance of the transition functions. 
\end{proof}
\end{lemma}

\textbf{Remark}. Generally one expects, in sufficiently non-pathological cases, that the existence of a $G$-equivariant ``connection'' on the fiber bundle is sufficient to define local trivializations satisfying the above properties (assuming the base space is path connected; otherwise we can just decompose the bundle into connected components.) We will not attempt to rigorously define what we mean by ``connection'' here.

\section{Equivalent characterizations of $\hat{G}$}
\label{appendix:structure_group_equivalence}
In this appendix, we will prove the equivalence of \eqref{eq:structure_group_1} and \eqref{eq:structure_group_2}.

Indeed, let $\Omega$ be a space with a transitive action of a continuous group $G$. We want to characterize the group of functions $f : \Omega \to \Omega$ that commute with this action (it will follow from the below results that any such function is a homeomorphism). First of all, if we fix some $\omega_* \in \Omega$, then by transitivity it is clear that the map $f$ is fully characterized by $f(\omega_*)$ since given any other $\omega \in \Omega$, we can find $g \in G$ such that $\omega = g \omega_*$, and then $f(\omega) = g f(\omega_*)$. The only thing that remains is to check the conditions such that (a) this gives a well-defined function of $f$, i.e.\ it is independent of the choice of $g$; and (b) it globally commutes with the $G$-action.

For (a): Suppose we have $g,g'$ such that $\omega = g \omega_* = g' \omega_*$. Then we have $(g')^{-1} g \in H$, where $H$ is the subgroup of $G$ that fixes $\omega_*$. Thus, $g = g'h$ for some $h \in h$. Then $g f(\omega_*) = g' f(\omega_*)$ is equivalent to $g'^{-1} g f(\omega_*) = f(\omega_*)$. Thus, we require that $f(\omega_*)$ is invariant under $H$. This is equivalent to saying that $g \in N_G(H)$, the normalizer of $H$ in $G$.

For (b): $f(g_0 \omega) = f(g_0 g \omega_*) = g_0 g f(\omega_*) = g_0 [ g f(\omega_*) ] = g_0 f(\omega)$, so global $G$-invariance is automatically satisfied given (a).

It follows that the group of maps $f$ that commute with the action of $G$ is isomorphic to $N_G(H)/H$ We can check that they satisfy the expected composition rule as follows. For $g_i \in N_G(H)$ with $i=1,2$, define $f_{g_iH}(\omega_*) = g_i \omega_*$. Then for a general $\omega = g \omega_*$, $f_{g_1H} \circ f_{g_2H}(\omega) = (g_1 h_1) (g_2 h_2) (g \omega_*) = g_1 g_2 g (h' \omega_*) = (g_1 g_2) (g \omega_*) = f_{g_1g_2H}(\omega)$. This is the expected result; it is independent of the choice of $h_1,h_2$.

\bibliography{refs_families}
\end{document}